\documentclass[twocolumn,aps,pra,superscriptaddress]{revtex4}

\usepackage{graphicx}
\usepackage{amsfonts}
\usepackage{amsmath}
\usepackage{amssymb}
\usepackage[colorlinks=true,citecolor=blue,urlcolor=blue]{hyperref}
\usepackage{natbib}

\begin{document}


\title{The thermodynamical cost of some interpretations of quantum theory. Reply to Prunkl and Timpson, and Davidsson}




\author{Ad\'an Cabello}
\email{adan@us.es}
\affiliation{Departamento de F\'{\i}sica Aplicada II, Universidad de Sevilla, E-41012 Sevilla, Spain}

\author{Mile Gu}
\affiliation{Centre for Quantum Information, Institute for Interdisciplinary Information Sciences, Tsinghua University, 100084 Beijing, China}
\affiliation{Centre for Quantum Technologies, National University of Singapore, 3 Science Drive 2, Singapore 117543, Singapore}
\affiliation{School of Physical and Mathematical Sciences, Nanyang Technological University, 21 Nanyang Link, 637371, Singapore}
\affiliation{Complexity Institute, Nanyang Technological University, 18 Nanyang Drive, 637723, Singapore}

\author{Otfried G\"uhne}
\affiliation{Naturwissenschaftlich-Technische Fakult\"at, Universit\"at Siegen,
Walter-Flex-Stra{\ss}e 3, D-57068 Siegen, Germany}

\author{Jan-{\AA}ke Larsson}
\affiliation{Institutionen f\"or Systemteknik, Link\"opings Universitet, SE-58183 Link\"oping, Sweden}

\author{Karoline Wiesner}
\affiliation{School of Mathematics, University of Bristol, University Walk, Bristol BS8 1TW, United Kingdom}


\date{\today}


\begin{abstract}
	Here we clarify the assumptions made and conclusions reached in our paper ``The thermodynamical cost of some interpretations of quantum theory'' [Phys.\ Rev.\ A \textbf{94}, 052127 (2016)], at the light of the criticisms of Prunkl and Timpson [Stud.\ Hist.\ Philos.\ Sci.\ Part B: Stud.\ Hist.\ Philos.\ Mod.\ Phys.\ \textbf{63}, 114 (2018)], and Davidsson (Master thesis, Stockholm University, 2018). We point out some misunderstandings and some weaknesses of the counterexample Prunkl and Timpson present to challenge our conclusion. We thus conclude, once more, that interpretations of quantum theory which consider the probabilities of measurement outcomes to be determined by objective properties of the measured system and satisfy the assumption that the measured system only has finite memory have a thermodynamical cost.
\end{abstract}


\maketitle


\section{Introduction}


In Ref.~\cite{PT18} Prunkl and Timpson, and in Ref.~\cite{Davidsson18} Davidsson, claim that the conclusion of our paper ``The thermodynamical cost of some interpretations of quantum theory'' \cite{CGGLW16} is incorrect. Here we will explain why we disagree with the arguments by these authors and clarify the result in \cite{CGGLW16} and its implications. 

Let us begin by summarizing the result in Ref.~\cite{CGGLW16}. Following Ref.~\cite{Cabello17}, we call {\em type~I interpretations of quantum theory} those in which the probabilities of measurement outcomes are determined by preexisting objective properties of the measured system. 
Bohmian mechanics \cite{Bohm52} and Everett's interpretation \cite{Everett57} are examples of type~I interpretations. In Bohmian mechanics, measurement outcomes are determined by the initial position of the measured system and by the state of the quantum potential which is a field permeating the whole universe. Similarly, in Everett's interpretation, measurement outcomes are determined by the universal wave function.

In contrast to that, {\em type~II interpretations} are those in which the probabilities of measurement outcomes are {\em not} determined by preexisting objective properties of the measured system. Quantum Bayesianism \cite{FS13} is an example of a type~II interpretation. A more comprehensive classification of the interpretations of quantum theory as type~I or type~II can be found in Ref.~\cite{Cabello17}.

The result in Ref.~\cite{CGGLW16} is that those type~I interpretations which {\em satisfy three additional assumptions} described below come with a thermodynamic cost. The proof is based on the following ideal experiment: A single quantum system (e.g., a trapped ion) is submitted to an unlimited sequence of measurements. Each of these measurements is randomly chosen from a given finite set. Ref.~\cite{LMZNCAH18} describes an experiment of this type: 53~million sequential dichotomic measurements, each of them randomly chosen from a set of 13 qutrit measurements. The three {\em additional} assumptions on which the argument is based are the following:

(i) The choice of measurement is random and independent of the system.

(ii) The system has limited memory. That is, the memory the system has access to is finite.

(iii) Landauer's principle \cite{Landauer61} holds: the erasure of one bit of information of the memory of the system must be accompanied by a corresponding entropy increase in its environment, causing the dissipation of, at least, $kT \ln 2$ units of heat.

Implicit is also the assumption that the system only takes information available in the past to produce the probabilities of the outcomes of future measurements. Our argument is as follows. In type~I interpretations satisfying assumptions (i)--(iii), the objective properties the system had previous to the measurement change during each measurement. In particular, the outcomes of future measurements change. Because of assumption~(i), the measured system does not know which particular sequence of measurements will be performed. Because of assumption~(ii), the system cannot store the outcomes for all possible sequences of measurements in its memory. Instead, after each measurement, the system has to add new information to its memory. The system can achieve this by, either, writing on empty memory or, if there is none, by {\em overwriting} part of its memory. {\em Overwriting constitutes a logically irreversible operation.} Since the memory of the system is finite, after a certain point the system must overwrite its memory. Then, because of Landauer's principle, the system must dissipate heat at least in an amount proportional to the information erased. 

If the set of possible measurements contains $n$ different choices, Ref.~\cite{CGGLW16} shows that the average dissipated heat per measurement performed {\em grows linearly with $n$}. Since, in principle, $n$ can be arbitrarily large, the system should dissipate an {\em unbounded} amount of heat in each measurement. This constitutes a kind of ultraviolet catastrophe or ``heat death'' \cite{Griffiths15}, which may lead to the conclusion that type~I interpretations satisfying assumptions (i)--(iii) are untenable and that thermodynamics constrains the possible interpretations of quantum theory.

In contrast, in type~II interpretations measurement outcomes are not governed or constrained by any preexistent property of the measured system. Measurement outcomes are simply manifestations of the fact that the ``system'' is preserved. Consequently, the system does not need to keep any information internally nor overwrite any preexisting information. Therefore, the system does not need to dissipate heat.


\section{``Central realist interpretations'' are not affected}


In Ref.~\cite{PT18}, Prunkl and Timpson write:
\begin{quote}
``[T]he authors [of Ref.~\cite{CGGLW16}] arrive at the conclusion that some quantum interpretations (including central realist interpretations) are associated with an excess heat cost and are thereby untenable. 
(\ldots)
If correct, this result would (\ldots)
	force us to abandon some of the most popular interpretations of quantum mechanics (Type~I interpretations include such favourites
	as de Broglie Bohm theory, Everett, and dynamical collapse theories
	such as GRW, for example).''
\end{quote}
This is simply not true, the conclusion of Ref.~\cite{CGGLW16} does not force us to abandon ``central realist interpretations'' such as Bohmian mechanics or Everett's interpretation because {\em they fail to satisfy assumption (ii)} as is clearly stated in Ref.~\cite{CGGLW16} when we point out that:
\begin{quote}
	''[O]ur result does not exclude Bohmian mechanics or the many worlds interpretation, since in both cases assumption (ii) is not satisfied (in Bohmian mechanics because the observed system includes an underlying continuous field and in the many worlds interpretation because the system itself splits in each measurement).''
\end{quote}
Instead, the message of Ref.~\cite{CGGLW16} is precisely that not only Bohmian mechanics and Everett's interpretation, but any type~I interpretation must allow the system to have access to infinite memory. 


\section{Prunkl and Timpson's ``counter-example''}


Prunkl and Timpson construct an example to illustrate their point that type~I interpretations do not imply that a quantum system dissipates heat when being measured. We see two problems with this example, each one of them disqualifying it as a ``counter''-example in our view: it is not an example of a type~I interpretation, and it does not do an accurate thermodynamic accounting. We now discuss these problems in turn. 


\subsection{Prunkl and Timpson's model is not type~I}
\label{subs:notanexample}


Firstly, in the framework in Ref.~\cite{CGGLW16}, the output of the measurement is important and cannot be ignored. To simulate the behavior of a quantum system, a well-defined input-output behavior needs to be simulated. This necessitates {\em a physical output}, irrespective of whether or not there is an external agent recording it. 
However, the model by Prunkl and Timpson does not output anything. Therefore, it fails to simulate the experiment in Ref.~\cite{CGGLW16}.

Secondly, measurements in the model by Prunkl and Timpson are described by partitioning the system of four states in two different ways. In order to account for a repeated outcome as predicted by quantum theory for the experiment in Ref.~\cite{CGGLW16}, the model by Prunkl and Timpson would require the partition to be {\em almost always} present. That is, performing a ``measurement'' would consist of:
\begin{itemize}
	\item[(A)]~Removing the existing partition, then
	\item[(B)]~rapidly inserting the partition corresponding to the new measurement choice (rapidly with respect to the free motion of the particle), then 
	\item[(C)]~finding out in which side of the partition the particle is, and then
	\item[(D)]~waiting for the particle position to be uniformly distributed in the region accessible to the particle. 
\end{itemize}
However, this does not give a type~I model. The epistemic state is the position of the particle {\em and} the chosen partition. There is no ontic state (i.e., no state that is responsible for the probabilities of the measurement outcomes). The mechanism Prunkl and Timpson propose always randomizes over time {\em relative to the previously performed measurement.} Therefore, ontic states before the measurement are not well-defined. Thus, one cannot talk about preexisting objective properties of the system determining the probabilities of the outcomes. Therefore, it is not a type~I model.
 
Moreover, it cannot be converted into a type~I model satisfying the limited memory assumption. It is possible to modify the model so it has well-defined ontic states. For that, {\em both} partitions need to be in place between the ``measurements'' which results in four regions. Then, a ``measurement'' would consist of: 
\begin{itemize}
	\item[(A')]~Removing {\em one} of the two partitions, then 
	\item[(B')]~finding out where the particle is, then 
	\item[(C')]~waiting for the position of the particle to be uniformly distributed in the region accessible to the particle, and then 
	\item[(D')]~re-inserting the removed partition.
\end{itemize}
This modified model has ontic states. However, the ontic state space is continuous and hence requires unbounded memory. Therefore, this modified model is type~I, but fails to satisfy the assumption of limited memory.


\subsection{Prunkl and Timpson's set-up does not consider a full thermodynamic cycle}


The second problem with the counter-example relates to the so-called RAND operation. The RAND operation was introduced \cite{Maroney05} as a proposed logically irreversible operation that does not generate heat following Landauer's principle, arguing that RAND randomizes a system state rather than resets it to a predefined state. It is therefore not a logical operation as the output is not deterministic. This was later discussed in Ref.~\cite{LPSG07}.

As seen in Fig.~2 of Ref.~\cite{PT18}, the system starts off, at time $t-1$, in a particular state ($x_{t-1} = 0, y_{t-1}=0$), then undergoes a RAND operation and is left, at time $t$, in one of three possible states (($x_{t} = 0, y_{t}=0$), ($x_{t} = 1, y_{t}=0$), or ($x_{t} = 1, y_{t}=1$)). In this case, the missing part of the cycle is the reset to the initial state $x_{t} = 0, y_{t}=0$. 

Let us examine how the ``reset'' operation would be done in this case: (a)~Remove the partition corresponding to the computational basis, (b)~use two pistons to compress each region in one direction to half of its initial size, then, (c)~insert the partition in question and return the pistons to their original position. This always outputs the epistemic state ``$|0\rangle$'' and has a work cost because of pressure against the compressing piston. 

Any other ``reset'' operation also must have a heat cost because otherwise, one could use it to construct a perpetuum mobile violating the first or second law of thermodynamics. One construction would be as follows:
(1)~``Measure'' in the computational basis, (2)~if the outcome is $0$, then stop, otherwise, ``measure'' in the phase basis, and repeat from (1). This procedure will eventually reset the system to $|0\rangle$ with no heat cost according to Prunkl and Timpson, since they claim their ``measurements'' have no heat cost. Such a reset operation could be followed with the first ``reset'' operation above reversed to extract work from the system in the state $|0\rangle$. This would clearly violate the second law of thermodynamics. Note that this is independent of the considerations in Subsection~\ref{subs:notanexample}, and applies even in the case where the system is allowed to reach thermal equilibrium between the two operations.


\subsection{A misconception of computational mechanics}


Another misconception concerns the information storage of the so-called causal states in the $\epsilon$-machine. Prunkl and Timpson \cite{PT18} write:
\begin{quote}
	``[\cite{CGGLW16}] say ``The average information that must be erased per measurement is the information contained in the causal state previous to the measurement, $S_{t-1}$, that is not contained in the causal state after the measurement, $S_t$.'' (Cabello et al., 2016, p.~2). This formulation is perhaps somewhat misleading, as it suggests that a particular causal state itself carries a certain amount of information. In fact, it is not {\em the} causal state to which we assign an entropy, but it is instead the probability distribution over causal states which is associated with an entropy and thereby with a Shannon information.''
\end{quote}
In our notation, ``the causal state previous to the measurement, $S_{t-1}$,'' is a random variable that has a probability distribution, the entropy of which, $H(S_{t-1})$, is the statistical complexity of the $\epsilon$-machine. The random variable $S_{t-1}$, conditioned on a particular value of the causal state at time $t$, $S_t = s_j$, is again a random variable with a probability distribution, but now a different one. The corresponding entropy is $H(S_{t-1}|S_t=s_j)$. This should then be averaged over all possible states $s_j$ to obtain the average information mentioned in the beginning of the quote above.


\section{Davidsson's ``refutation''}


Davidsson's Master thesis \cite{Davidsson18} contains, in our opinion, a well-written overview and formal treatment of Landauer's principle, focusing on extending the principle to quantum systems. There is also a rederivation of our bound that confirms the correctness of our calculations. Perhaps it is appropriate to point out that our bound $I_{\rm erased}(n)>n$ is exact, and not approximate as claimed in \cite{Davidsson18} (there denoted $\langle I_{\varepsilon}\rangle>n$), since
\begin{equation}
\begin{split}
\sum_{j=0}^{2^{n+1}-1}&\cos^2(\tfrac{\pi j}{2^{n+1}})=\sum_{j=0}^{2^{n}-1}\cos^2(\tfrac{\pi j}{2^{n+1}})+\sin^2(\tfrac{\pi j}{2^{n+1}})=2^n,
\end{split} 
\end{equation}
so that
\begin{equation}
 \label{eq:9}
\begin{split}
I_{\rm erased}(n)&=-\sum_{j=0}^{2^{n+1}-1}\frac{\cos^2\left(\frac{\pi j}{2^{n+1}}\right)}{2^n}
\log \frac{\cos^2\left(\frac{\pi
j}{2^{n+1}}\right)}{2^n}\\
&>-\sum_{j=0}^{2^{n+1}-1}\frac{\cos^2\left(\frac{\pi j}{2^{n+1}}\right)}{2^n}
\log \frac{1}{2^n}=n.
\end{split} 
\tag{\cite{CGGLW16}:9}
\end{equation}

Davidsson continues with an argument that the entropy of the quantum system under study should be calculated from the quantum description. The argument reviews the differences between the classical and quantum treatments and continues:
\begin{quote}
In this framework, \textit{von Neumann entropy} $S_N$ (\ldots) can be motivated as the basis for calculating entropies in physical quantum systems, and thus becomes the basis for generalizing the \textit{Landauer bound} and \textit{Landauer’s principle} (\ldots) to quantum systems \ldots\ 
Thus we should calculate entropy in a quantum system from the density matrix, according to equation (\cite{Davidsson18}:216), and the calculation made by Cabello et al.\ appear[s] conceptually inaccurate. \ldots\ 
In conclusion, it is premise \cite{Davidsson18}:7.3 [Landauer's principle holds] that fails. Not because Landauer’s principle does not hold, but because it does not apply to decreases of entropy associated with causal states in this particular set-up.
\end{quote}
At this point, we must again stress that we address type~I interpretations of quantum theory, where the probabilities of measurement outcomes are determined by preexisting objective properties of the measured system. 
Then, we are allowed to \emph{calculate entropy from these preexisting objective properties} (which includes use of causal states in one-to-one correspondence with nonorthogonal quantum states), and use the output as a lower bound.
There is no conceptual inaccuracy here: the preexisting objective properties, together with assumptions (i)--(iii), enables the corresponding increased lower bound.

That being said, we find it sensible to use Davidsson's generalized Landauer's principle for other interpretations, i.e., type~I interpretations where assumptions (i)--(iii) do not hold, and type~II interpretations. In this case,
\begin{quote}
This implies that entropy in this qubit-system is, not only bounded from below by 0, but also bounded from above by 1 bit. 
Clearly, the maximum change in entropy is therefore bounded by 1 bit, implying that \ldots\ any lower bound cannot go above $kT\ln 2$.
\end{quote}
This underlines the difference between the possible bounds from type~I interpretations under assumptions (i)--(iii), and other interpretations.

\null

\section{Conclusion}


The conclusion of Ref.~\cite{CGGLW16} is not that type~I interpretations such as Bohmian mechanics or Everett's are untenable, as Prunkl and Timpson \cite{PT18} claim, but rather that in all type~I interpretations the measured system must have access to infinite memory. Secondly, the counter-example that Prunkl and Timpson provide is not an example of a type~I interpretation since it does not provide preexisting objective properties that determine the probabilities of measurement outcomes. Even if we attempt to modify it to contain such properties, it will use infinite memory to provide the outcome probabilities. Furthermore, the system properties suggested by Prunkl and Timpson can be used to construct a perpetuum mobile, i.e., these properties violate the first or second law of thermodynamics. This should conclude any argument on the issue.

Davidsson's thesis \cite{Davidsson18} contains a generalization of the Landauer principle to the quantum case, and we do not dispute general use of the resulting lower bound. But for type~I interpretations we maintain that the preexisting objective properties that determine outcome probabilites, together with assumption (i)--(iii), enable the increased lower bound of Ref.~\cite{CGGLW16}.


\section*{Acknowledgements}


We thank C.\ E.\ A.\ Prunkl, C.\ Timpson, and E.\ Davidsson for bringing their works to our attention. AC thanks the FQXi Large Grant ``The Observer Observed: A Bayesian Route to the Reconstruction of Quantum Theory,'' the Spanish MICINN Project No.\ FIS2017-89609-P with FEDER funds, and the Knut and Alice Wallenberg Foundation. MG thanks the National Research Foundation of Singapore Fellowship No.\ NRF-NRFF2016-02, the National Research
Foundation and L'Agence Nationale de la Recherche joint Project No.\ NRF2017-NRF-ANR004 VanQuTe,
and the Singapore Ministry of Education Tier 1 RG190/17. OG thanks the ERC (Consolidator Grant No.\ 683107/TempoQ).


\end{document}